\documentclass[twocolumn,pra,floatfix,citeautoscript,nofootinbib,superscriptaddress]{revtex4}
\usepackage{amsbsy}
\usepackage{latexsym,epsfig,graphicx}
\usepackage{dcolumn}
\usepackage{graphicx}
\usepackage{subfigure}
\usepackage{comment}
\usepackage{color}
\usepackage{bm}
\usepackage{mathrsfs}
\usepackage{amsfonts}
\usepackage{amsmath}
\usepackage{color}
\usepackage{amssymb}
\usepackage{xspace}
\usepackage{epstopdf}
\usepackage{tabularx}
\usepackage{longtable}
\usepackage[colorlinks=true, letterpaper=true, pdfstartview=FitV, linkcolor=blue, citecolor=blue, urlcolor=blue]{hyperref}
\usepackage[normalem]{ulem}

\pdfoutput=1
\input{tcilatex}
\begin{document}

\title{Higher-order topological corner states induced solely by onsite
potentials with mirror symmetry}
\author{Ya-Jie Wu}
\thanks{wuyajie@xatu.edu.cn}
\affiliation{School of Sciences, Xi'an Technological University, Xi'an 710032, China}
\author{Wen He}
\affiliation{School of Sciences, Xi'an Technological University, Xi'an 710032, China}
\author{Ning Li}
\affiliation{School of Sciences, Xi'an Technological University, Xi'an 710032, China}
\author{Zhitong Li}
\affiliation{School of Science, Beijing University of Posts and Telecommunications, Beijing 100876, China}
\affiliation{State Key Laboratory of Information Photonics and Optical Communications, Beijing University of Posts and Telecommunications, Beijing 100876, China}
\author{Junpeng Hou}
\thanks{ryanhou@fb.com}
\affiliation{Department of Physics, The University of Texas at Dallas, Richardson, Texas 75080, USA}

\begin{abstract}
Higher-order topological insulators have triggered great interests because of
exhibitions of non-trivial bulk topology on lower-dimensional boundaries like corners and hinges. While such
interesting phases have been investigated in a plethora of systems by tuning
staggered tunneling strength or manipulating existing topological phases, here we
show that a higher-order topological phase can be driven solely by mirror-symmetric
onsite potentials. We first introduce a simple chain model in one dimension that mimics
the Su-Schrieffer-Heeger-like model. However, due to the lack of internal symmetries
like chiral or particle-hole symmetry, the energies of the topological edge
modes are not pinned at zero. Once the model is generalized to two dimensions, we
observe the emergence of topological corner modes. These corner modes are
intrinsic manifestation of non-trivial bulk band topology protected by mirror
symmetry, and thus, they are robust against symmetry-preserved perturbations.
Our study provides a concise proposal for realizing a class of higher-order
topological insulators, which involves only tuning onsite energies. This can be
easily accessible in experiments and provides a different playground for
engineering topological corner modes.
\end{abstract}

\maketitle

\section{Introduction}

Since the discovery of quantized multipole insulators, higher-order (HO)
topological phases and materials have attracted great interests because of
their novel bulk-boundary correspondences \cite%
{Benalcazar2017,Song2017,Langbehn2017,Benalcazar2018,Franca2018,hli2020}. In contrast to
conventional (or first-order) topological phases, the topologically protected
boundary states of HO topological phase exhibit lower dimensions.
In other words, $r$th-order (co-dimension $r\geqslant 2$) topological phases in $d$
dimensions host $\left( d-r\right)$-dimensional localized states,
rather than $d-1$ dimensional edge states. For example, in two-dimensional ($2$D)
second-order topological insulators, the boundary states manifest as zero-dimensional ($0$D) corner states. A variety of candidates have been proposed to host
HO topological phases \cite%
{Ezawah2018,Ezawat2018,Ezawa2018,FLiu2019,HFan2019,HXue2019,Peng2019,
Zangeneh2019,Park2019,Pozo2019,Sheng2019,RChen2020,Qi2020,Huang2020,
Chen2020,cbhua2020,Kirsch2021,Schindler2018,Xie2018,ChenX2019,Hassan2019,Qiao2020,CALi2020,Agarwala2020,yijiawu2020,jhwang2021,Bliu2021}.

Spatial symmetries enrich topological phases from an aspect
differing from conventionally internal symmetries (i.e., particle-hole,
time-reversal and chiral symmetries) \cite{Fu2007}. These topological
crystalline phases have been classified in a unified framework \cite%
{Shiozaki2014}. Later, crystalline symmetries have also been
shown to play an important role in different types of HO topological
insulators and superconductors \cite%
{Miert2018,Khalaf2018,Schindler2018,Benalcazar2019,Cornfeld2019,Cornfeld2021}.
While there have been different approaches for realizing HO topological
phases, most of them are driven from an existing non-trivial first-order
topological phase or rely on extending some models similar
to Su-Schrieffer-Heeger (SSH) model by adjusting staggered hopping.
More recently, a handful of studies have shown that both first-order and HO
topological insulators can be driven by tuning non-Hermitian effects including
on-site gain and loss rates \cite{Parto2018,Luo2019,Wu2020}.
While non-Hermiticity has brought interesting
aspects into topological phases of matter, they are harder to engineer and
control in general. Thus, an interesting question naturally arises: whether existing first-order topological phase and non-Hermitian
effects are necessary for driving HO topological phases?

In this paper, we answer the above question by enriching the family of HO
topological insulators with a class of HO topological phases induced
solely by real and mirror-symmetric onsite potentials. We first introduce a one-dimensional ($1$D) chain with mirror symmetry and show that a topological phase can be driven
through only onsite energy difference between sublattices. Contrary to the usual SSH model
and its derivatives, our model is protected by mirror symmetry, instead of chiral
symmetry, and thus, the edge modes have non-zero energies. We then generalize
the $1$D model to a Wannier-type second-order topological insulator, which is
driven from a Dirac semimetal by only onsite potentials. Such a topologically non-trivial
phase can be characterized by Wannier centers. To show the flexibility
of the proposed platform, we further present an intrinsic second-order topological insulator
on a square lattice and characterize its topology using edge polarizations and
quadrupole moment. Finally, to confirm the topological protection of this class of second-order
topological insulators, we impose mirror-symmetry-preserved perturbations and show that the
corner modes are robust up to a global energy shift.

This paper is organized as follows. In Sec. \ref{S1}, we introduce a $1$D
lattice with inversion symmetric onsite potentials, which serves as the base
for our HO topological models. In Sec. \ref{S2}, we start with a honeycomb lattice with
mirror-symmetric onsite potentials, and explore the HO topological phases.
In Sec. \ref{S3}, we turn to the study of a square lattice, and introduce edge
polarizations and quadrupole moment to characterize the corresponding HO topological phase. Conclusions and discussions are drawn in Sec. \ref{S4}.

\section{1D superlattice with mirror-symmetric potentials} \label{S1}
\begin{figure}[tbp]
\centering\includegraphics[width=0.48\textwidth]{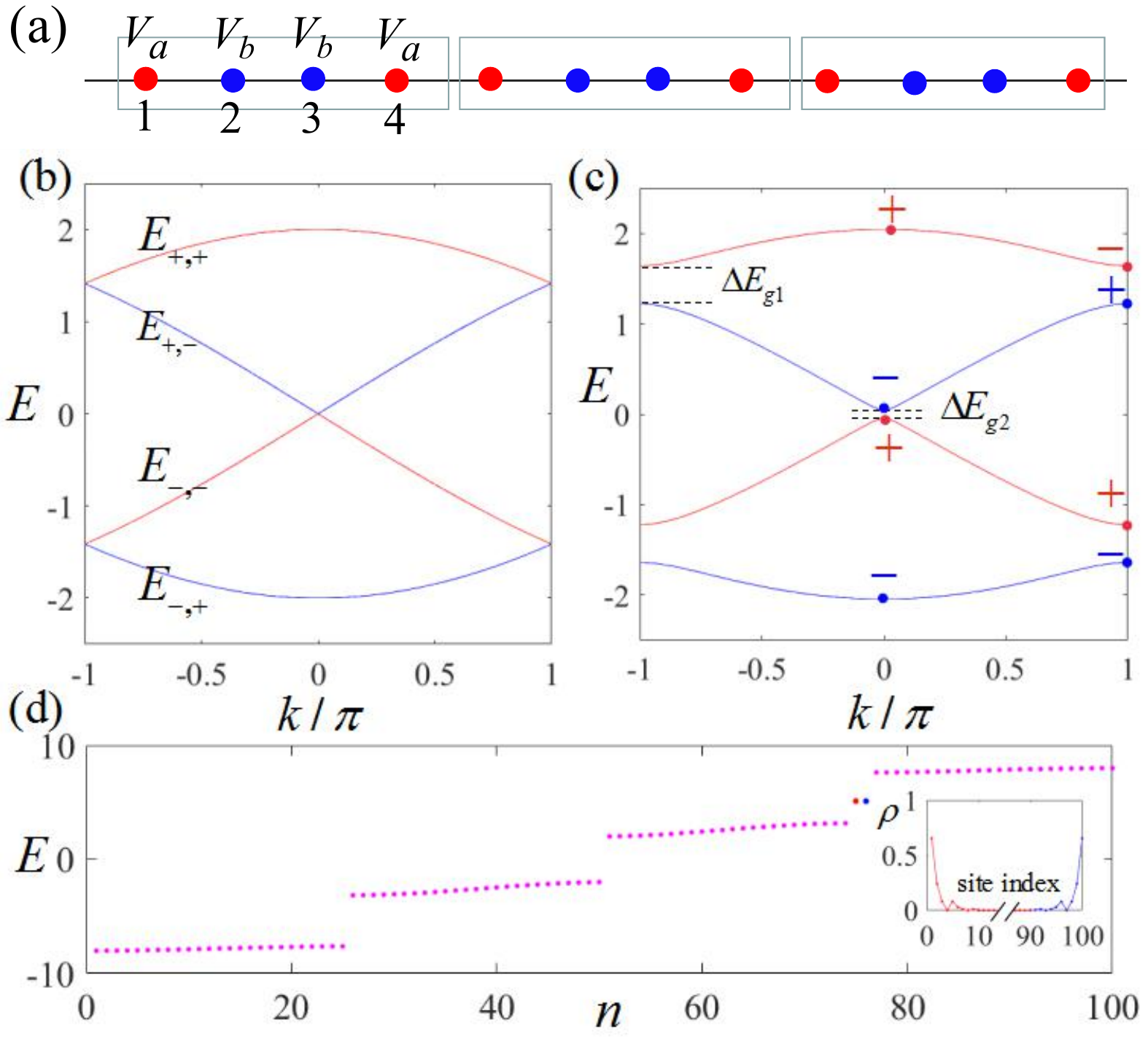}
\caption{(a) Illustration of 1D lattice with onsite potentials $V_{a}$ and $V_{b}$. (b) and (c) Energy spectra at different onsite potentials with $%
V=0$ and $V=0.3$. (d) Eigenenergies for 1D chain with $100$ sites and $V=4.0$ under open-boundary conditions. $n$ denotes eigenstate index. The inset
showcases the particle density versus site index for states indicated by blue and red dots in (d).}
\label{1D}
\end{figure}

For simplicity, we consider a 1D superlattice consisting of two sublattice sites with
different onsite potentials $V_a$ and $V_b$, while inter- and intra-site coupling $t$ is
uniform. A minimal non-trivial model with
mirror (inversion) symmetry has a unit cell $(V_a,V_b,V_a)$. Here for better demonstration, we use 
a configuration of $(V_a,V_b,V_b,V_a)$, as shown in Fig. \ref{1D} (a), and the corresponding system Hamiltonian in momentum
space reads
\begin{eqnarray}
h\left( k\right) &=& \frac{t}{2}\left(1+\cos k\right)\sigma _{x}\tau _{x}+\frac{t}{%
2}\left(1-\cos k\right)\sigma _{y}\tau _{y}+t\sigma _{0}\tau _{x}\nonumber\\
&&+\frac{t}{2}\sin k\left(\sigma _{y}\tau _{x}+\sigma _{x}\tau
_{y}\right)+V_{-}\sigma _{z}\tau _{z}+V_{+}\sigma _{0}\tau _{0} 
\end{eqnarray}%
under the basis $\hat{\psi}_{k}=\left( \hat{a}_{1,k},\hat{a}_{2,k},\hat{a}%
_{3,k},\hat{a}_{4,k}\right) ^{T}$, where $V_{\mp }=\left( V_{a}\mp
V_{b}\right) /2$, $\sigma _{0}$ and $\tau _{0}$ are identity matrices. The
Hamiltonian $h\left( k\right) $ preserves mirror symmetry with $\mathcal{M}%
h\left( k\right) \mathcal{M}^{-1}=h\left( -k\right) $, where $\mathcal{M}%
=\sigma _{x}\tau _{x}$. In the following, we set $V_{+}=0$, i.e., $V_{a}=-V_{b}=V>0$ without loss of
generality since $V_{+}I$ only shifts the energy bands globally, but doesn't
change the topological number of each bands because the eigenvectors remain
invariant. The four energy bands are given by%
\begin{eqnarray} \nonumber
E_{\pm ,+} &=&\pm \sqrt{2t^{2}+V^{2}+\sqrt{2t^{2}\left(
t^{2}+2V^{2}+t^{2}\cos k\right) }}, \\ \nonumber
E_{\pm ,-} &=&\pm \sqrt{2t^{2}+V^{2}-\sqrt{2t^{2}\left(
t^{2}+2V^{2}+t^{2}\cos k\right) }}.
\end{eqnarray}%
When $V=0$, there are four gapless energy bands as shown in Fig. \ref%
{1D} (b). The bands $E_{+,+}$ $\left( E_{-,+}\right) $ and $E_{+,-}$ $\left(
E_{-,-}\right) $ touch at momentum point $k=\pi $, and the bands $E_{+,-}$
and $E_{-,-}$ touch at $k=0$ with linear dispersions. As $V$ increases, the
bands $E_{+,+}$ $\left( E_{-,+}\right) $ and $E_{+,-}$ $\left(
E_{-,-}\right) $ are separated with an energy gap $\Delta E_{g1}=\sqrt{%
t^{2}+\left( t+V\right) ^{2}}-\sqrt{t^{2}+\left( t-V\right) ^{2}}$ at $k=\pi 
$, and there opens an energy gap $\Delta E_{g2}=\sqrt{t^{2}+V^{2}}-t$
between $E_{+,-}$ and $E_{-,-}$ at $k=0$ (see Fig. \ref{1D} (c)).

Since the 1D superlattice preserves mirror symmetry, each energy band would
contribute a quantized topological invariant. We define the topological
invariant as 
\begin{equation}
\eta =-\frac{1}{\pi }\doint \mathcal{A}_{k}dk_{x},  \label{Equ1}
\end{equation}%
with $\mathcal{A}_{k}=-i\left\langle u_{m,k}|\partial
_{k_{x}}u_{n,k}\right\rangle $. After calculations, we find that $\eta
=0,0,1,-1$ for bands $E_{-,+}$, $E_{-,-}$, $E_{+,-}$ and $E_{+,+}$,
respectively. The topological invariants can also be equivalently defined by 
$\eta =-\left[ m\left( 0\right) -m\left( \pi \right) \right] /2$, where $%
m\left( 0\right) $ and $m\left( \pi \right) $ are parities at momentum
points $k=0$ and $\pi $, respectively. In Fig. \ref{1D} (d), we denote even
and odd parity as "$+$" and "$-$", respectively. We can also obtain
consistent topological invariants for each band with that from Eq. (\ref{Equ1})
.

To present the bulk-boundary correspondence, we calculate energies for a
lattice under open boundary conditions. The energy level distributions are
shown in Fig. \ref{1D} (d). It shows that two states emerge in the
gap between the third band and fourth band from bottom to top. The inset of
Fig. \ref{1D} (d) presents the particle density for these two states, which
shows that these two in-gap states are localized at two ends of lattices,
similar to the celebrated SSH model. This behavior confirms the
bulk-boundary correspondence for this topological system.

When $V<0$, the topological invariants for bands $E_{-,+}$, $E_{-,-}$, $%
E_{+,-}$ and $E_{+,+}$ would be changed to $1,-1,0,0$, and two
localized states emerge in the gap between the first and second band
from bottom to top. In this sense $V=0$ is a critical topological phase
transition point for each energy band.

\section{2D honeycomb lattice with mirror-symmetric potentials}
\label{S2} 
\begin{figure}[tbp]
\centering\includegraphics[width=0.48\textwidth]{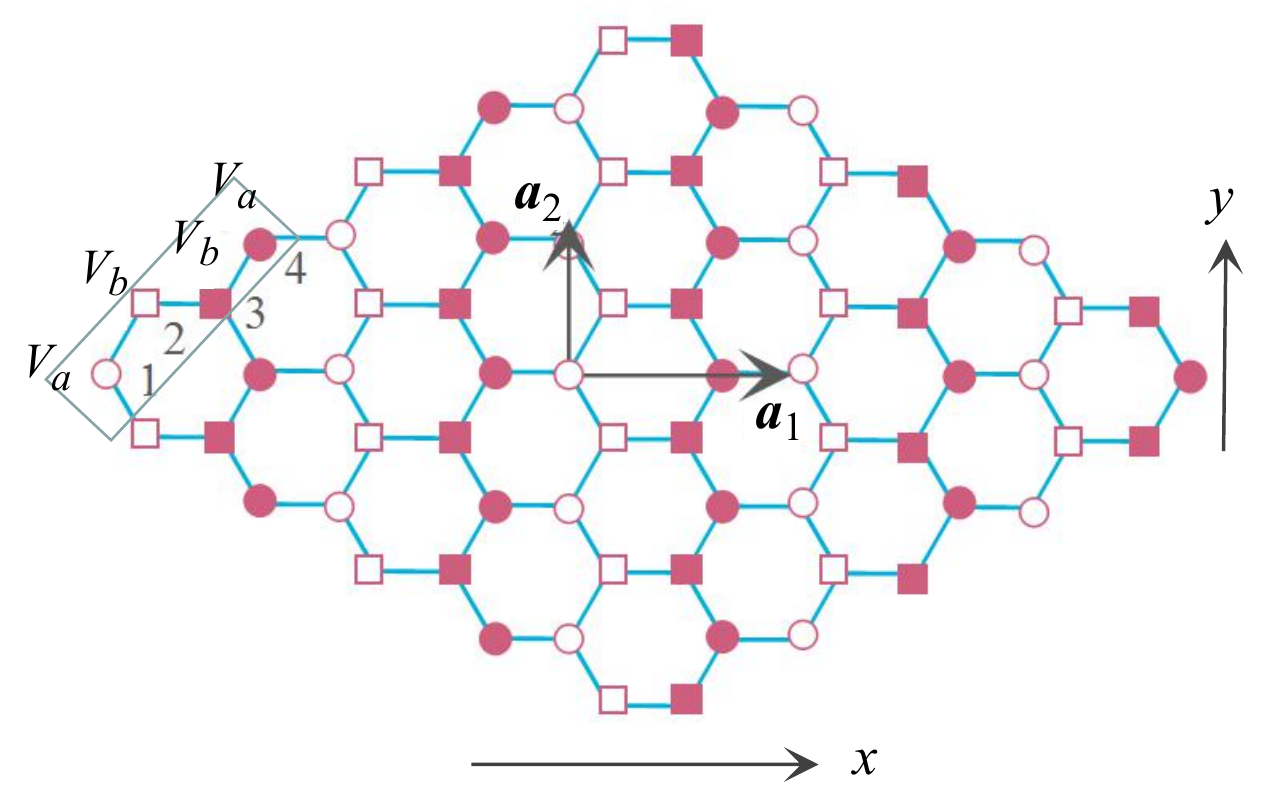}
\caption{Illustration of a honeycomb lattice with mirror-symmetric potentials along $x$. Each unit-cell consists of four sublattice sites indexed by $1-4$.}
\label{honeycomb}
\end{figure}
We consider a graphene lattice with mirror-symmetric onsite
potentials $(V_{a},V_{b},V_{b},V_{a})$ along $x$, as shown in Fig. \ref{honeycomb}. The single-particle
Hamiltonian is written as $\hat{H}=\hat{H}_{0}+\hat{H}_{\mathrm{p}}$. The
first term on r.h.s. reads $\hat{H}_{0}=-t\sum_{\left\langle
i_{m},j_{n}\right\rangle }\left( \hat{a}_{i_{m}}^{\dagger }\hat{a}%
_{j_{n}}+h.c.\right) $ with $t$ the coupling between nearest-neighbor sites $%
\left\langle i_{m},j_{n}\right\rangle $, and the mirror-symmetric potential
is described by $\hat{H}_{\mathrm{p}}=\sum_{m,i}V_{m}\hat{a}_{i_{m}}^{\dagger }\hat{a}%
_{i_{m}}$ with $V_{m=1,4}=V_{a}$ and $V _{m=2,3}=V_{b}$ representing
the mirror symmetric onsite potentials. The operator $\hat{a}%
_{i_{m}}^{\dagger }$ $\left( \hat{a}_{i_{m}}\right) $ creates
(annihilates) a mode at site $i_{m}$. The mirror-symmetric onsite potentials
enlarge the unit-cell of the usual honeycomb lattice. Each unit-cell consists
of four sublattices. The Bravais vectors are now described by 
$\bm{\mathit{a}}_{1}=\left( 3,0\right) $ and $\bm{\mathit{a}}_{2}=\left( 0,\sqrt{3}\right) $, as shown in Fig.  \ref{honeycomb}. The
first Brillouin zone (BZ) decreases correspondingly. Here we have set the lattice
spacing of the honeycomb lattice to be unit. The total Hamiltonian $H$ in
momentum space can be written as $\hat{H}=\sum_{k}\hat{\psi}_{k}^{\dagger
}h\left( k\right) \hat{\psi}_{k}$ under the basis $\hat{\psi}_{k}=\left( 
\hat{a}_{1,k},\hat{a}_{2,k},\hat{a}_{3,k},\hat{a}_{4,k}\right) ^{T}$ with
\begin{eqnarray}
h\left( k\right) &=&\alpha _{k}\sigma _{0}\tau _{x}-t\sin k_{a_{2}}\sigma
_{0}\tau _{y}+\beta _{k}\sigma _{x}\tau _{x}+\gamma _{k}\sigma _{y}\tau _{y}
\notag \\
&&+\frac{t}{2}\sin k_{a_{3}}\left( \sigma _{x}\tau _{y}+\sigma _{y}\tau
_{x}\right) +V_{-}\sigma _{z}\tau _{z}+V_{+}I, 
\end{eqnarray}%
where $\alpha _{k}=t\left[ 1+\cos \left( \sqrt{3}k_{y}\right) \right] $, $%
\beta _{k}=\frac{1}{2}t\left( 1+\cos k_{a_{3}}\right) $, $\gamma _{k}=\frac{1%
}{2}t\left( 1-\cos k_{a_{3}}\right) $, $k_{a_{2}}=\sqrt{3}k_{y}$, $k_{a_{3}}=%
\bm{\mathit{k}}.\left(\bm{\mathit{a}}_{1}+\bm{\mathit{a}}_{2}\right) =3k_{x}+\sqrt{3}k_{y}$, and $I=\sigma _{0}\tau _{0}$.

When $V_{a}=V_{b}$, the Hamiltonian $\hat{H}$ preserves inversion symmetry $%
\mathcal{P=}\sigma _{x}\tau _{x}\mathcal{U}$, time-reversal symmetry $%
\mathcal{T=KU}$ and $C_{3}$ rotation symmetry, where $\mathcal{U}=diag\left(
e^{ikb_{1}-ik\frac{b_{3}}{2}},e^{-ik\frac{b_{3}}{2}},e^{ik\frac{b_{3}}{2}%
},e^{-ikb_{1}+ik\frac{b_{3}}{2}}\right) $ is a diagonal unitary matrix and $%
\mathcal{K}$ is a conjugation operator. These symmetries lead to the system
hosting two locally and globally stable Dirac points with geometric phases $%
\pm \pi $.

When $V_{a}\neq V_{b}$, $C_{3}$ symmetry would be broken and
Dirac points may be gapped. However, in this case the model also preserves
mirror symmetries, i.e., $\mathcal{M}_{x/y}h\left( k_{x/y}\right) \mathcal{M}%
_{x/y}^{-1}=h\left( -k_{x/y}\right) $ with $\mathcal{M}_{x}=\sigma _{x}\tau
_{x}\mathcal{U}$ and $\mathcal{M}_{y}=\sigma _{0}\tau _{0}\mathcal{U}$.
Without loss of generality, we again set $V_{a}=-$ $V_{b}=V$ in the following.
The four energy bands in momentum space are solved as
\begin{eqnarray} \notag
E_{\pm ,+}\left( k\right)  &=&\pm \sqrt{V^{2}+t^{2}\left( 3+2\cos \sqrt{3}%
k_{y}\right) +2\sqrt{\alpha _{k}}}, \\ \notag
E_{\pm ,-}\left( k\right)  &=&\pm \sqrt{V^{2}+t^{2}\left( 3+2\cos \sqrt{3}%
k_{y}\right) -2\sqrt{\alpha _{k}}},
\end{eqnarray}%
where $\alpha _{k}=t^{2}\left( t^{2}+V^{2}+t^{2}\beta _{k}\right) $ with $%
\beta _{k}=\cos 3k_{x}+\left( 1+\cos 3k_{x}\right) \cos \sqrt{3}k_{y}$. 

Compared to the conventional graphene model with $V=0$, the energy bands are
folded and Dirac points shift to $\bm{\mathit{K}}=\left( 0,2\sqrt{3}\pi /9\right) 
$ and $\bm{\mathit{K}}^{\prime }=\left( 0,-2\sqrt{3}\pi /9\right) $ as plotted in Fig. \ref%
{2Dspectra}(a$_{1}$). As the potential $V$ increases, two Dirac points
remain massless while they approach each other in
momentum space because the local stability is protected by $%
\mathcal{P}$ and $\mathcal{T}$ symmetries. They merge at the
time-reversal-invariant point $\left( 0,\pi /\sqrt{3}\right) $ at $V=V_{c1}=t
$, as shown in Fig. \ref{2Dspectra}(b$_{1}$). If $V$ increases further,
an energy gap opens as shown in Fig. \ref{2Dspectra} (c$_{1}$). In the
following, we will showcase the topological nature for each energy band.

\subsection{Topological bands and topological invariants}

From above symmetry analysis, both $\mathcal{PT}$ symmetry and mirror
symmetry are respected along $x$ for the system. The mirror symmetry $%
\mathcal{M}_{x}$ guarantees the non-trivial quantization of polarization\
along the $x$ direction. To present the polarization as the bulk property,
we construct a Wilson loop operator $\mathcal{W}_{x,k}$ in the $x$
direction, where $k$ represents the base point of the loop. We define the
Bloch wave function of the occupied energy bands with negative energies as $%
\left\vert u_{m,k}\right\rangle $, where $H\left( k\right) \left\vert
u_{m,k}\right\rangle =E_{m}\left( k\right) \left\vert u_{m,k}\right\rangle $
with normalization condition $\left\langle u_{m,k}|u_{n,k^{\prime
}}\right\rangle =\delta _{m,n}\delta _{k,k^{\prime }}$. The Wilson loop
operator is described by $W_{x,k}=F_{x,k+N_{x}\Delta k_{x}}...F_{x,k+\Delta
k_{x}}F_{x,k}$, where the elements for $F_{x,k}$ are defined by $\left[
F_{x,k}\right] ^{m,n}=\left\langle u_{m,k+\Delta k_{x}}|u_{n,k}\right\rangle 
$ with $\Delta k_{x}=2\pi /N_{x}$ and $N_{x}$ the number of unit cells in
the $x$ direction. The topological invariant at each $k_{y}$ is then defined by $\eta
_{x}\left( k_{y}\right) =-\frac{i}{\pi }\mathrm{Tr}\left( \ln W_{x,k}\right) 
$, which forms the Wannier band. It is quantized under mirror symmetries. In
the thermodynamic limit, the topological invariant $\eta _{x}\left(
k_{y}\right) $ is given by%
\begin{equation}
\eta _{x}\left( k_{y}\right) =-\frac{1}{\pi }\mathrm{Tr}\left( \doint 
\mathcal{A}_{k}dk_{x}\right) ,
\end{equation}%
where ${A}_{k}$ is a non-Abelian Berry connection
with $\left( \mathcal{A}_{k}\right) _{mn}=-i\left\langle u_{m,k}|\partial
_{k_{x}}u_{n,k}\right\rangle $. Following similar steps, the topological
invariant $\eta_{y}\left( k_{x}\right) $ could be obtained at each $k_{x}$.
Finally, the topological invariant, namely the Wannier center of Wannier
bands is defined as $\left( \eta _{x}^{\prime },\eta _{y}^{\prime }\right) $
with $\eta _{x/y}^{\prime }=\frac{1}{4b_{y/x}}\doint\eta \left(
k_{y/x}\right) dk_{y/x}$, where the reciprocal vectors are $b_{x}=\pi /3$
and $b_{y}=\pi /\sqrt{3}$.

\begin{figure}[tbp]
\centering\includegraphics[width=0.48\textwidth]{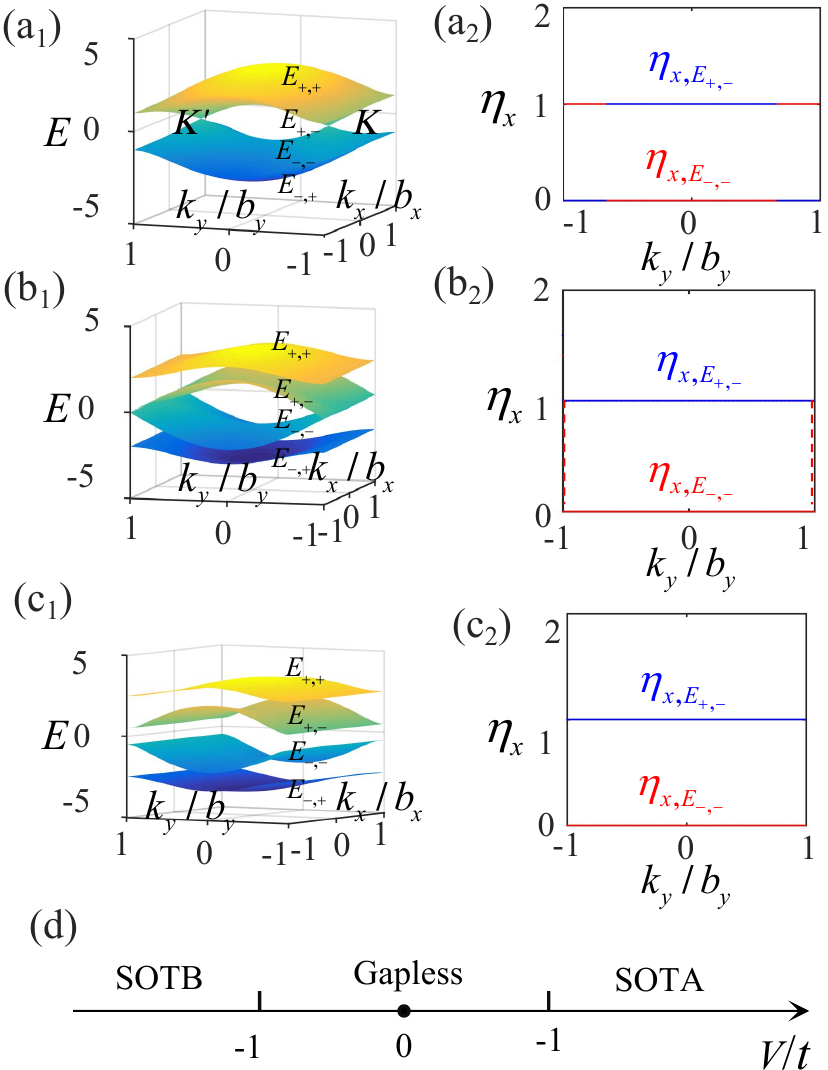}
\caption{(a$_{\text{1}}$)-(c$_{\text{1}}$) Energy spectra at different
onsite potentials. (a$_{\text{2}}$)-(c$_{\text{2}}$) Topological invariants
for the energy band $E_{-,-}$ (red lines) and the energy band $E_{+,-}$
(blue lines) corresponding to (a$_{\text{1}}$)-(c$_{\text{1}}$). The
strength for each case is (a$_{\text{1}}$) and (a$_{\text{2}}$) $V=0$, (b$_{%
\text{1}}$) and (b$_{\text{2}}$) $V=1.0$, (c$_{\text{1}}$) and (c$_{\text{2}}
$) $V=1.5$. (d) Phase diagram versus $V/t$. Other parameters are set to be $%
t=1$, $b_{x}=\protect\pi /3$, and $b_{y}=\protect\pi /\protect\sqrt{3}$.}
\label{2Dspectra}
\end{figure}

When $V=0$, $\hat{H}$ becomes a conventional graphene model. The topological
invariant for the band $E_{-,-}$ is $\eta _{x,-,-}\left( k_{y}\right) =1$ if 
$-\pi /\sqrt{3}<k_{y}<-2\sqrt{3}\pi /9$ or $2\sqrt{3}\pi /9<k_{y}<\pi /\sqrt{%
3}$ and $\eta _{x,E_{-,-}}\left( k_{y}\right) =0$ otherwise. The topological
invariant for the band $E_{+,-}$ is $\eta _{x,E_{+,-}}\left( k_{y}\right) =0$
if $-\pi /\sqrt{3}<k_{y}<-2\sqrt{3}\pi /9$ or $2\sqrt{3}\pi /9<k_{y}<\pi /%
\sqrt{3}$ and $\eta _{x,E_{+,-}}\left( k_{y}\right) =1$ otherwise. The
topological invariants for the bands $E_{+,+}$ and $E_{-,-}$ are $\eta
_{x,E_{-,-}}\left( k_{y}\right) =0$ for any $k_{y}$, as plotted in Fig. \ref%
{2Dspectra}(a$_{2}$). For $V=V_{c_{1}}=t$, the energy gap closes at momentum
lines $k=\left( k_{x},\pi /\sqrt{3}\right) $ with any $k_{x}$, as presented in Fig. \ref{2Dspectra} (b$_{1}$). The topological invariants for each band has been shown in Fig. \ref
{2Dspectra} (b$_{1}$), where $\eta _{x,E_{+,-}}\left( \pm \pi /\sqrt{3}%
\right) $ and $\eta _{x,E_{-,-}}\left( \pm \pi /\sqrt{3}\right) $ are not
well defined indicated by the dashed lines. When $V>V_{c_{1}}$, the energy gap
opens between bands $E_{+,-}$ and $E_{-,-}$. We find $\eta
_{x,E_{+,+}}\left( k_{y}\right) =-1$, $\eta _{x,E_{+,-}}\left( k_{y}\right)
=1,\eta _{x,E_{-,-}}\left( k_{y}\right) =0,$ and $\eta _{x,E_{-,+}}\left(
k_{y}\right) =0$ for any $k_{y}$ [see Fig.~\ref{2Dspectra}(c$_{2}$)].
Namely, when $V>V_{c_{1}}$, the Wannier centers of the bands $E_{+,-}$, $E_{-,-}$ and $E_{-,+}$ are $\eta
_{y,E_{+,-}}^{\prime }=1/2$, $\eta _{y,E_{-,-}}^{\prime }=0,\eta
_{y,E_{-,+}}^{\prime }=0$,
respectively. 

Therefore, if $V>V_{c_{1}}$, the total Wannier center of the lowest three Wannier bands is
quantized to a non-trivial value $\left( 1/2,0\right) $ with lowest three energy bands ($E_{+,-}$, $E_{-,-}$ and $%
E_{-,+}$) being occupied. We dubbed this phase as second-order topological
phase A (SOTA). Similar cases happen when $V<$ $V_{c_{2}}=-t$, where we observe
that the bands $E_{+,+}$ and $E_{+,-}$ become topologically trivial, while
bands $E_{-,-}$ and $E_{-,+}$ are topologically non-trivial. These lead to
that the Wannier center becomes $\left( 1/2,0\right) $ if the bands $E_{-,-}
$ is occupied. This phase is referred to second-order topological phase B
(SOTB). In summary, the topological phase diagram is shown in Fig.~\ref%
{2Dspectra}(d).

Here, we would like to point out that the quantization of topological
invariant $\eta _{x}^{\prime }$ is guaranteed by mirror symmetry along the $%
x $ direction. It is robust against weak mirror-symmetric perturbations as
long as the corresponding energy gap doesn't close.

\label{Sec3}

\subsection{Topological corner modes and interface modes}

\begin{figure}[h]
\centering\includegraphics[width=0.48\textwidth]{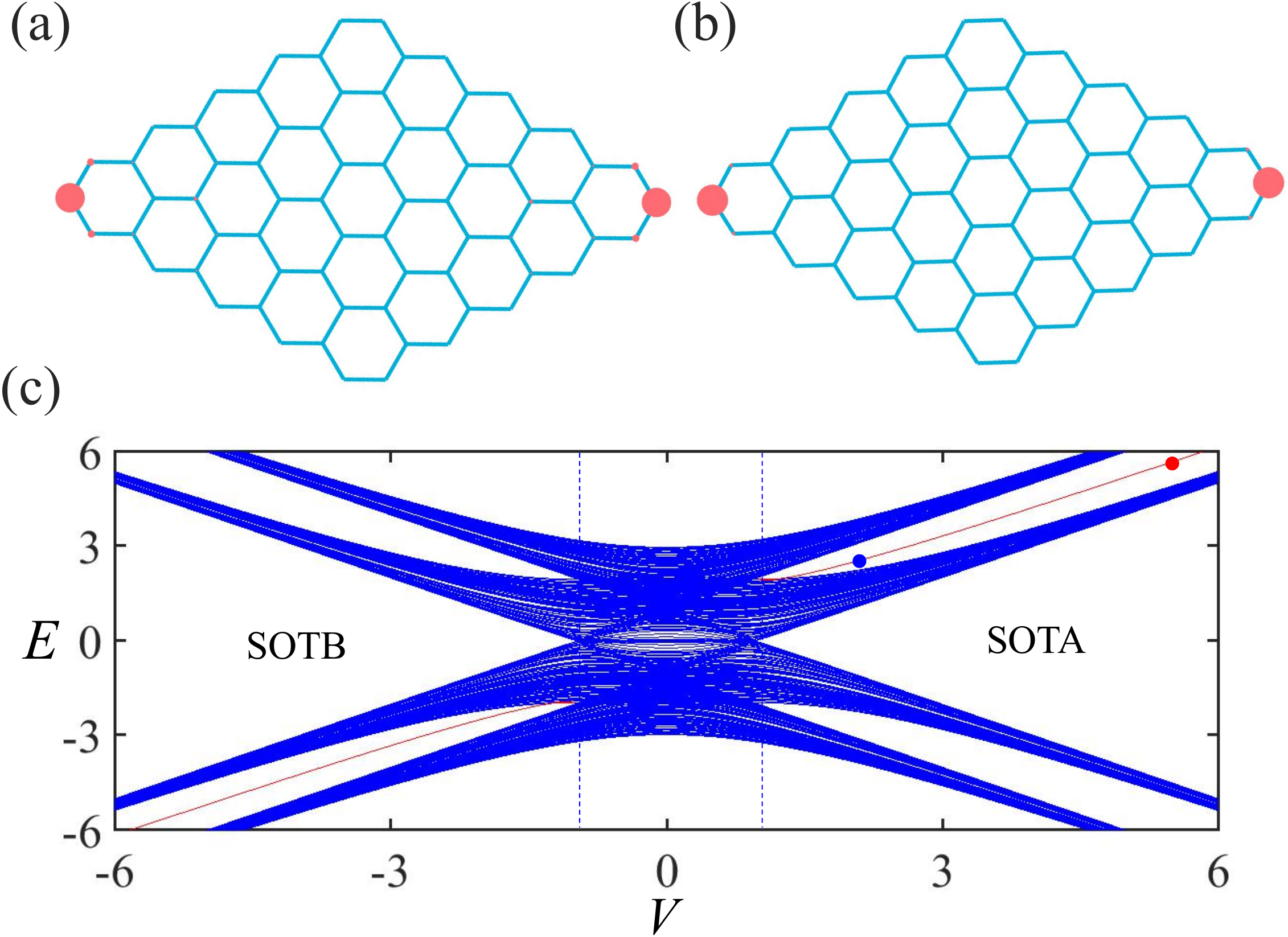}
\caption{(a) Spatial density distribution of the corner modes with $V=2$
indicated by blue dot in (c) and the radii of the pink disk is proportional
to local density. (b) Similar to (a) but plotted with a different onsite
potential $V=5.5$ indicated by red dot in (c), demonstrating the corner
modes become more localized as the strength of onsite potential increases.
(c) Eigenspectrum versus potential $V$. The red lines denote two-fold
degenerate corner modes.  Common parameter is set to be $t=1$.}
\label{Corner}
\end{figure}

Consider a sample shown in Fig. \ref{Corner}(a). We set the
parameter $V>V_{c_{1}}$ so that the system is in
a topological phase SOTA with the Wannier center quantized to $\left( 1/2,0\right) $%
. First take the case $V=2.0$ as an example and its numeric results are
plotted in Fig. \ref{Corner}(a) and (c). Two degenerate
modes indicated by the blue dots emerge in the energy gap as shown in Fig. %
\ref{Corner}(c). The corresponding particle density distributions have been
shown in Fig. \ref{Corner}(a). It presents that the two degenerate modes are
localized at two horizontal corners of the given sample. Fig. \ref{Corner}(b) shows the case $V=5.5$, in which the corner modes become more localized as the strength of potential increases. Similarly, we observe
that the topological corner modes also exist when $V<V_{c_{2}}$ in
topological phase SOTB. 

So far, we have focused on the special case with $V_{a}=-V_{b}$. We remarked
that the quantization of a non-trivial topological invariant (Wannier center)
is guaranteed by the mirror symmetry. It means that it is also respected when 
$V_{a}\neq V_{b}$ since $V_{+}=\left( V_{a}+V_{b}\right) /2$ only shifts
energies globally and $V_{-}=\left( V_{a}-V_{b}\right) /2$ determines the
wave functions. Therefore, it is expected that the system is non-trivial when 
$\left\vert V_{-}\right\vert >V_{c_{1}}$, as long as the energy gap between
nearby bands remains open. Here, we also would like to remark
that while the mirror-symmetric potential perturbations may shift energies
of corner modes, the Wannier center of the system is invariant, and the
corner modes remain localized. This is different from conventional
HO topological system where the energies of corner modes are usually pinned at
zero.

\begin{figure}[tbp]
\centering\includegraphics[width=0.48\textwidth]{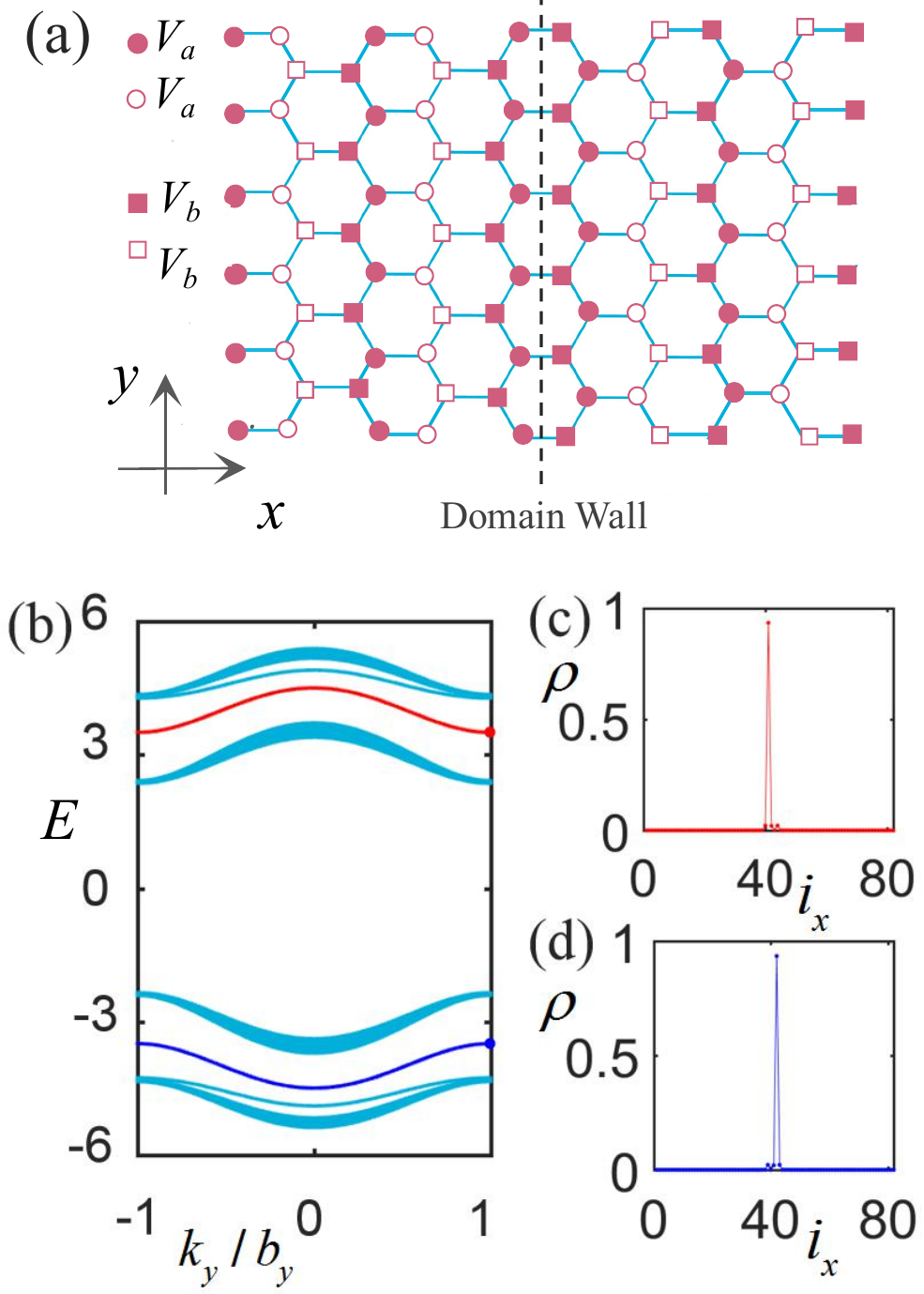}
\caption{(a) Two graphene sheets with a domain wall in
between, which is highlighted by the dashed line. As they possess opposite
topological invariants, topological interface modes near the domain wall naturally arise
. (b) Energy spectra versus $k_{y}$ for two graphene sheets with a domain wall in between and an open-boundary condition
along $x$ (total $N_{x}=82$ sites). The blue and red curves indicate the
interface modes. (c) and (d) depict the the density distribution of two
localized interface states. We set parameters $t=1$ and $V=3.2$.}
\label{domain}
\end{figure}

As discussed above, a graphene model with appropriate mirror-symmetric
potentials $V$ is a HO topological insulator characterized by
Wannier center $\left( 1/2,0\right) $. In the following, we
consider two graphene sheets separated by a domain wall as sketched in Fig. %
\ref{domain}(a). Here the translation symmetry of graphene lattice is broken
along $x$ direction, but the translation symmetry is respected along $y$. In
the following, we take $k_{y}$ as a system parameter and treat $\hat{H}%
(k_{y})$ as a quasi-one dimensional chain. The Hamiltonian $\hat{H}$ is then
written as 
\begin{equation}
\hat{H}=\sum_{k_{y}}\hat{H}\left( k_{y}\right) =\sum_{k_{y}}\hat{H}%
_{0}\left( k_{y}\right) +\hat{H}_{\mathrm{p}}\left( k_{y}\right) ,
\end{equation}%
with%
\begin{eqnarray} \notag
\hat{H}_{0}\left( k_{y}\right) &=&-t\sum\nolimits_{i_{x}}\hat{a}%
_{1,i_{x},k_{y}}^{\dagger }\hat{a}_{4,i_{x}-1,k_{y}}+\epsilon _{k_{y}}\hat{a}%
_{1,i_{x},k_{y}}^{\dagger }\hat{a}_{2,i_{x},k_{y}} \\ \notag
&&+\hat{a}_{2,i_{x},k_{y}}^{\dagger }\hat{a}_{3,i_{x},k_{y}}+\epsilon
_{k_{y}}^{\ast }\hat{a}_{3,i_{x},k_{y}}^{\dagger }\hat{a}%
_{4,i_{x},k_{y}}+h.c., \\
H_{\mathrm{p}}\left( k_{y}\right) &=&\sum_{m,i_{x}}V_{m}\gamma
_{m}a_{m,i_{x},k_{y}}^{\dagger }a_{m,i_{x},k_{y}}
\end{eqnarray}%
where $\epsilon _{k_{y}}=t\left( 1+e^{i\sqrt{3}k_{y}}\right) $ and the
domain wall structure is given by $\gamma _{m=1,4}=-\gamma _{m=2,3}=1$
(left-hand side of the domain wall) and $\gamma _{m=1,4}=-\gamma _{m=2,3}=-1$
(right-hand side), as depicted in Fig. \ref{domain}(a).

The energy spectra of the system can be derived from $\hat{H}\left(
k_{y}\right) \left\vert u\left( k_{y}\right) \right\rangle $ $=E\left(
k_{y}\right) \left\vert u\left( k_{y}\right) \right\rangle $. We set
the strength of appropriate potentials so that both graphene sheets are in
different topological phases, i.e., the left and right ones are in SOTA and
SOTB, respectively. Through numerical calculations, we obtain energy
spectra of graphene sheets as shown in Fig. \ref{domain} (b), and
observe localized states at the interface, i.e., the states with positive
and negative energies localizing at the left-hand and right-hand side of the domain
wall, respectively, as plotted in Fig. \ref{domain} (c) and (d). This confirms that
two different topological phases (SOTA and SOTB) indeed exhibit different topological properties and such a setup can be used as
the experimental setup of photonic higher-order topological insulators in
graphene lattices.

For a domain-wall structure, we may also consider the general case $%
(V_{a},V_{b},V_{b},V_{a})$ with $V_{a}\neq -V_{b}$. The numeric calculations
also demonstrate the existence of topological localized interface modes.
Finally, we also consider mirror-symmetric perturbations and find that,
although the energies of localized states and bulk states vary, the
topological interface states always localize at the domain wall. In this
sense, the topological interface states are robust and mirror-symmetry protected.

\section{2D square lattice with mirror-symmetric potentials}
\label{S3} 
\begin{figure}[tbp]
\centering\includegraphics[width=0.48\textwidth]{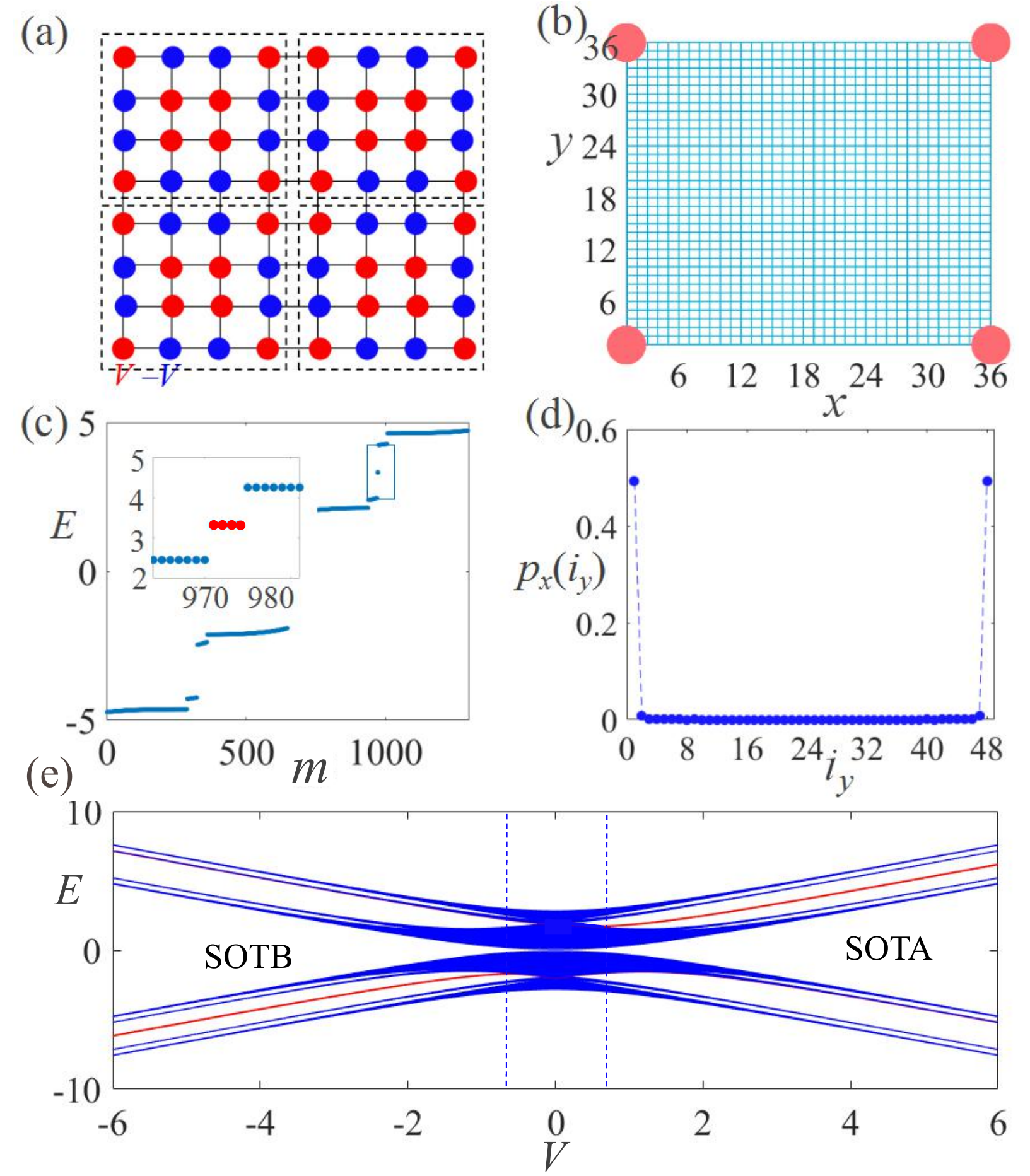}
\caption{(a) Illustration of a square lattice with mirror-symmetric potentials. Each unit-cell consists of sixteen sublattice sites. (b) Spatial density distribution of the corner modes indicated in (c). The
radii of spots is proportional to the particle density. (c) Eigenenergy level
distributions for a $36\times36$ lattice. The inset shows enlarge portion of rectangle in (c), where four red dots indicate corner modes. (d) Edge polarizations for a configurations with open boundary along $x$
but periodic boundary along $y$. Common parameters are $t=1$, $V=3$ in (b)-(d). 
(e) Eigenenergies versus potential $V$. The red lines in SOTA and SOTB represent four-fold degenerate corner modes.}
\label{squ}
\end{figure}

We now consider a square lattice with mirror-symmetric potentials and each square plaquette enclosing a $\pi $ flux, as shown in Fig. \ref{squ} (a). The Hamiltonian is written as%
\begin{eqnarray}
h\left( k\right) &=&t\sigma _{+}^{x}\sigma _{-}^{x}\sigma _{0}^{y}\sigma
_{0}^{y}+t\sigma _{0}^{x}\sigma _{+}^{x}\sigma _{0}^{y}\sigma
_{0}^{y}+te^{-ik_{x}}\sigma _{+}^{x}\sigma _{+}^{x}\sigma _{0}^{y}\sigma
_{0}^{y}  \notag \\
&&+t\sigma _{0}^{x}\sigma _{z}^{x}\sigma _{+}^{y}\sigma _{-}^{y}+t\sigma
_{0}^{x}\sigma _{z}^{x}\sigma _{0}^{y}\sigma _{+}^{y}+te^{-ik_{y}}\sigma
_{0}^{x}\sigma _{z}^{x}\sigma _{+}^{y}\sigma _{+}^{y}  \notag \\
&&+h.c.+V\sigma _{z}^{x}\sigma _{z}^{x}\sigma _{z}^{y}\sigma _{z}^{y},
\end{eqnarray}%
where $t$ and $V$ denote the coupling between nearest-neighbor sites and onsite potential, respectively. $\mathbf{\sigma }^{x}$ and $\mathbf{\sigma }^{y}$ are Pauli matrices acting on the degrees of freedom spanned along $x$ and $y$, respectively,
while $\sigma _{0}^{x}$ and $\sigma _{0}^{y}$ are identity matrices. The
ladder operator $\sigma_{\pm }^{x}$ reads $%
\sigma _{\pm }^{x}=\left( \sigma _{x}^{x}\pm i\sigma _{y}^{x}\right) /2$
and $\sigma _{\pm }^{y}$ is defined similarly.
The Hamiltonian preserves mirror symmetries along both $x$ and $y$
as $\mathcal{M}_{x}^{^{\prime }}H\left( k_{x},k_{y}\right) 
\mathcal{M}_{x}^{^{\prime }-1}=h\left( -k_{x},k_{y}\right) $ and $\mathcal{M}%
_{y}^{^{\prime }}h\left( k_{x},k_{y}\right) \mathcal{M}_{y}^{^{\prime
}-1}=h\left( k_{x},-k_{y}\right) $, where $\mathcal{M}_{x}^{^{\prime }}=\sigma_{x}^{x}\sigma_{x}^{x}\sigma_{0}^{y}\sigma_{z}^{y}$ and $\mathcal{M}_{y}^{^{\prime }}=\sigma_{0}^{x}\sigma_{0}^{x}\sigma_{x}^{y}\sigma_{x}^{y}$. There are eight pairs of energy bands and each
pair is doubly degenerate. When $V=0$, the two central energy bands touch
at Dirac point $\Gamma =\left( 0,0\right) $. When $V \neq 0$, an energy
gap opens at $\Gamma $ with the gap $\Delta E_{g}=2\left( \sqrt{V^{2}+2t^{2}}%
-\sqrt{2}t\right) $. In addition, there also opens a gap between the first-two pairs and second-two pairs of bands form
top to bottom as potential $V$ increases, which implies that the topological phase
transition may occur with opening the gap.

Consider a square sample with $36\times 36$ sites with $V=3.0$. We compute
its eigenenergy level distributions as shown in Fig. \ref{squ}(c). It showcases
that there are four energy modes in the energy gap. After plotting
the particle density distributions (see Fig. \ref{squ}(b)), we find these
modes are localized at four corners of the sample.

To characterize the topological properties of corner states, we compute the
edge polarizations $p_{x}^{\mathrm{edge},y}$ ($p_{y}^{\mathrm{edge},x}$)
using Wilson loops on a torus geometry where the lattice has open boundary
along $y$ ($x$) but periodic boundary along $x$ ($y$). The polarization distribution
along $y$ is defined by \cite{Benalcazar2017,Benalcazar2018} 
\begin{equation}
p_{x}\left( i_{y}\right) =\frac{1}{N_{x}}\sum_{j,k_{x},\alpha ,n}\left\vert %
\left[ u_{k_{x}}^{n}\right] ^{i_{y},\alpha }\left[ v_{k_{x}}^{j}\right]
^{n}\right\vert ^{2}\nu _{x}^{j}.
\end{equation}%
Here $N_{x}$ is the number of unit-cell along the $x$ direction. $\left[
u_{k_{x}}^{n}\right] ^{i_{y},\alpha }$ denotes the ($i_{y},\alpha $)-th
component of occupied state $\left\vert u_{k_{x}}^{n}\right\rangle $, where $%
i_{y}$ and $\alpha $ are the site index along $y$ and sublattice degrees of
freedom along $x$, respectively. $\left[ v_{k_{x}}^{j}\right] ^{n}$ is the $n$th
component of $j$th eigenvector corresponding to the Wannier values $\nu
_{x}^{j}$ of the Wannier Hamiltonian $H_{\mathcal{W}_{x}}=-i\ln \mathcal{W}%
_{x}$. $\mathcal{W}_{x}$ represents the Wilson loop operator, i.e., $%
\mathcal{W}_{x}=F_{x,k_{x}+\left( N_{x}-1\right) \Delta
k_{x}}...F_{x,k_{x}+\Delta k_{x}}F_{x,k_{x}}$ with $\left[ F_{x,k_{x}}\right]
^{mn}=\left\langle u_{k_{x}+\Delta k_{x}}^{m}|u_{k_{x}}^{n}\right\rangle $
and $\Delta k_{x}=2\pi /N_{x}$. 

Figure \ref{squ}(d) presents the
edge polarization $p_{x}\left( i_{y}\right) $ versus site index $i_{y}$ with $V=3$ and the occupied state number $n_{occ}=4n_{y}*3/4$, where $n_{y}$ is the site number along $y$. 
Similar results are obtained on a torus geometry where the lattice has open boundary along $x$ but periodic boundary along $y$.
These results implies that localized modes exist at the corners of the sample, consistent with the results in Fig. \ref{squ}(b) and (d). 

In addition, this topological phase can also be characterized by quadrupole
moment \cite{Agarwala2020}. We consider $N_{o}$ states are occupied represented as $\left\vert
\chi _{l}\right\rangle =\sum_{j\alpha }\phi _{l,j\alpha }\left\vert j\alpha
\right\rangle $, where $l=1,...,N_{o}$, and $j\alpha $ denotes indices of
unit-cells and sublattice sites with $j=1,...,N_{u}$ and $\alpha =1,2,...,16$. By
arranging $N_{o}$ occupied state columnwise, we construct a unitary matrix $%
U $ with dimension $16N_{u}\times N_{o}$. The density of particles is defined
by 
\begin{equation}
n=-\frac{i}{2\pi }\text{Tr}\ln U^{\dagger }OU.
\end{equation}%
Here $O$ is a diagonal $16N_{u}\times 16N_{u}$ dimensional matrix with
elements $O_{4\left( j-1\right) +\alpha_{x} ,4\left( j-1\right) +\alpha_{y} }=\exp
\left( i2\pi x_{j\alpha_{x} }y_{j\alpha_{y} }/N_{x}N_{y}\right) $, where $N_{x}$ and 
$N_{y}$ are unit-cell numbers along $x$ and $y$, and $\alpha_{x},\alpha_{y}=1,2,3,4$. To characterize the
topological properties, we subtract the contribution of density in the
atomic limit, represented by $n_{at}=n_{f}\sum_{j\alpha_{x}\alpha_{y} }x_{j\alpha_{x}
}y_{j\alpha_{y} }/\left( N_{x}N_{y}\right) $ with $n_{f}=N_{o}/\left(
16N_{u}\right) $, and define quadrupole moment as $Q_{xy}=n-n_{at}$ $%
\left( \func{mod}1\right) $. Through numeric calculations with periodic
boundary conditions, we obtain $Q_{xy}=1/2$ with $N_{o}=16N_{u}*3/4$ when $V=3$, indicating a
second-order topological phase emerges driven by mirror-symmetric potentials. We dubbed this phase as SOTA  phase, as shown in Fig. \ref{squ} (e).

Finally, we plot eigenenergies versus potential $V$, and observe topological corner modes indicated by red lines with four-fold degenerates in both SOTA and SOTB phases, as shown in Fig. \ref{squ} (e). We would like to remark that this HO phase is robust against weak mirror-symmetric perturbations while the energies of corner modes may globally shift (see Appendix for details).

\section{Conclusion and Discussion}
\label{S4} 
\begin{figure}[tbp]
\centering\includegraphics[width=0.48\textwidth]{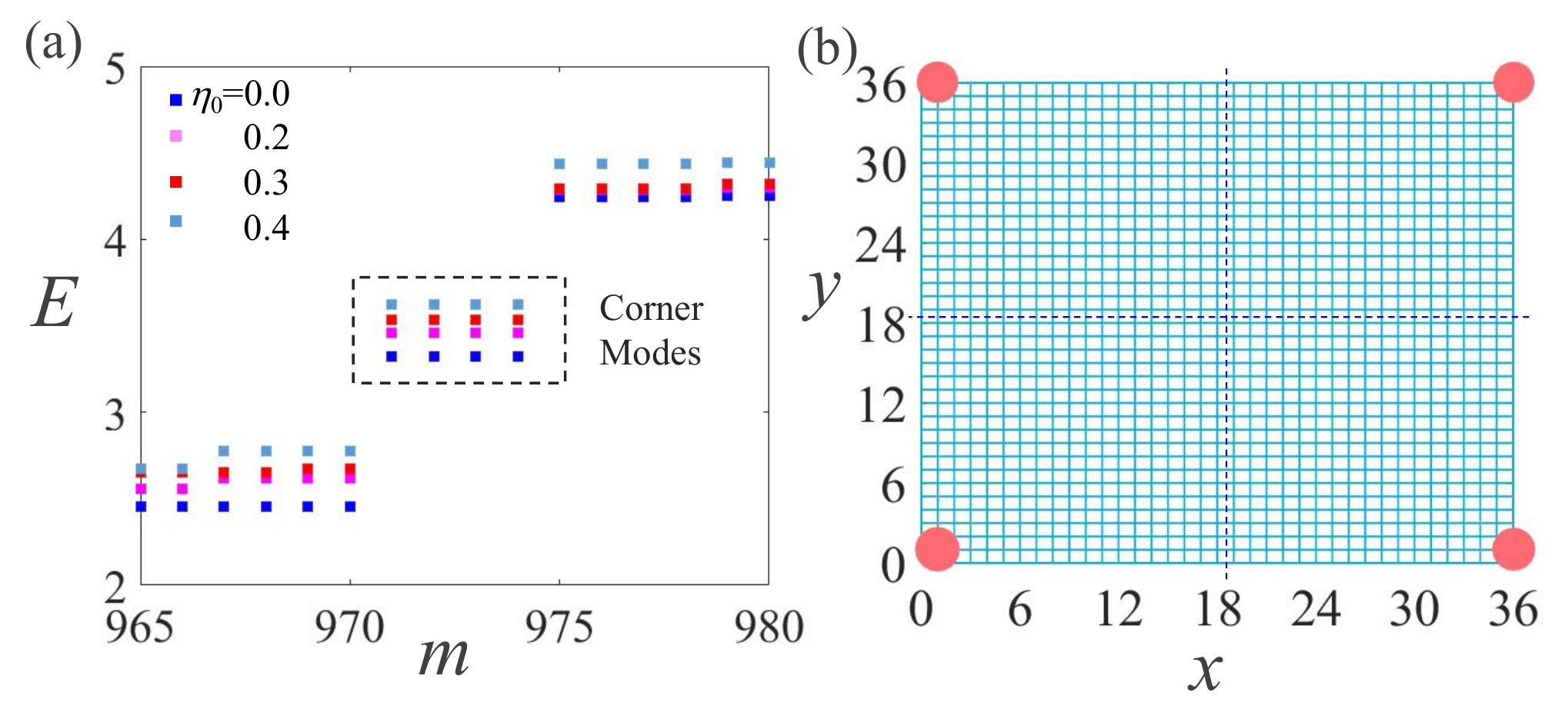}
\caption{(a) Eigenenergies $E$ versus the state index $m$ in the presence of
mirror-symmetric random potentials with different amplitude $\protect\eta %
_{0}$. (b) The particle density distributions of in-gap modes indicated by the
red square in dashed box in (a) with $\protect\eta _{0}=0.3$. Blue dashed
lines represent two mirrors along $x$ and $y$. Common parameter is set to be $t=1$.}
\label{random}
\end{figure}

In contrast to most of previous proposals to implement
higher-order topological phases, we introduce an interesting and accessible
platform for engineering and controlling HO topological phases by
solely tuning real onsite potentials.
The topological corner
modes are in the band energy gap and localize at 0D boundaries. In the presence of
mirror-symmetric disorders, this HO topological phase is robust in the sense that the
in-gap localized modes remain at corners and topological invariant doesn't
change, while the energies of in-gap corner modes shift since this model
doesn't preserve chiral symmetry or particle-hole symmetry. Our proposed
model can be readily implemented in a range of systems like cold atoms, optics and acoustics.
The proposed models also provides a playground for studying novel topological
phenomena with, for example, non-Hermitian effects or nonlinear interactions.

In summary, we have studied HO topological phases induced solely
by mirror-symmetric onsite potentials. The model can also be
generalized to implementing third-order topological phases in three
dimensions. Because manipulating onsite energies are accessible in most experimental platforms and our proposal does not require fine tuning of hopping strength, our scheme
provides a realistic playground for the experimental study of HO
topological phases in both quantum and classical systems.

\begin{acknowledgments}
This work is supported by the Scientific Research Program Funded by the
Natural Science Basic Research Plan in the Shaanxi Province of China
(Programs No. 2021JM-421 and No. 2019JM-001), the NSFC under the Grant No.
11504285, the Scientific Research Program Funded by Shaanxi Provincial
Education Department under the grant No. 18JK0397, and the Young Talent fund
of the University Association for Science and Technology in Shaanxi, China
(Program No. 20170608).
\end{acknowledgments}

\appendix

\section{Robustness of corner states against perturbations}

\label{app} We consider two cases to show the robustness of corner states
against perturbations. First we impose mirror-symmetric perturbation on
onsite potentials described as $\hat{H}_{ap}=\sum_{i}\eta _{i}\hat{a}%
_{i}^{\dagger }\hat{a}_{i}$, where $\eta _{i=\left( i_{x},i_{y}\right)
}=\eta _{0}\kappa _{i_{x,}i_{y}}$, $\eta _{0}$ is the amplitude of the
random potential, and $\kappa _{i_{x,}i_{y}}=\kappa
_{N_{x}-i_{x}+1,N_{y}-i_{y}+1}\in \left[ 0,1\right] $ is a random quantity
for $i_{x}\leq N_{x}/2$ and $i_{y}\leq N_{y}/2$ (see Fig. \ref{random}(b)). We take a $36\times 36$
square lattice as an example and numerically compute its energies, as shown
in Fig. \ref{random}(a). We observe that the energies for four in-gap states
acquire a finite energy shift. We showcase the density distribution of these
four states in Fig. \ref{random}(b). It shows that they remain localized at
four corners of the sample, which indicates the induced second-order
topological phase with corner modes are robust.

We next consider the\ general random perturbation on onsite potentials
represented as $H_{bp}=\sum_{i}\delta _{i}a_{i}^{\dagger }a_{i}$, where $%
\delta _{i}=\delta _{0}\kappa _{i}$, $\kappa _{i}\in \left[ 0,1\right] $ is
a random number. Through numeric calculations we find four ingap states may
acquire different energies. It implies the corner states no longer have the
same energies in the absence of mirror symmetries. However, these modes
remain localized at corners as long as they are in the band-gap.

To summarize, the corner states are robust against mirror-symmetric
perturbations, but they acquire same non-zero energies.

\end{document}